\documentclass[10pt,a4paper,twocolumn,superscriptaddress]{revtex4}

\usepackage{textcomp}
\usepackage{color}
\usepackage{amsmath}
\usepackage{mathrsfs}
\usepackage{graphicx}
\usepackage{subfig}
\usepackage{epsfig}
\usepackage{physics}
\usepackage{amssymb}
\usepackage{hyperref}
\usepackage{todonotes}
\usepackage[final]{pdfpages}

\newcommand{\DL}{\mathrm{DL}}
\newcommand{\SL}{\mathrm{SL}}
\newcommand{\tauRbar}{\overline{\tau_R}}
\newcommand{\dbar}{d\hspace*{-0.08em}\bar{}\hspace*{0.1em}}

\usepackage[squaren,Gray]{SIunits}

\usepackage{pgfplots}
\pgfplotsset{compat=newest}
\usepgfplotslibrary{groupplots}
\usepgfplotslibrary{polar}
\usepgfplotslibrary{smithchart}
\usepgfplotslibrary{statistics}
\usepgfplotslibrary{dateplot}
\usepgfplotslibrary{ternary}

\begin{document}

\title{Manipulation and control of temporal cavity solitons with trapping potentials}

\author{Nicolas Englebert}
\affiliation{Service OPERA-\textit{Photonique}, Universit\'e libre de Bruxelles (U.L.B.), 50~Avenue F. D. Roosevelt, CP 194/5, B-1050 Brussels, Belgium}
\affiliation{Department of Electrical Engineering, California Institute of Technology, Pasadena, California 91125, USA}
\author{Corentin Simon}
\affiliation{Service OPERA-\textit{Photonique}, Universit\'e libre de Bruxelles (U.L.B.), 50~Avenue F. D. Roosevelt, CP 194/5, B-1050 Brussels, Belgium}
\author{Carlos Mas Arabí}
\affiliation{Service OPERA-\textit{Photonique}, Universit\'e libre de Bruxelles (U.L.B.), 50~Avenue F. D. Roosevelt, CP 194/5, B-1050 Brussels, Belgium}
\affiliation{Institut Universitari de Matem\`{a}tica Pura i Aplicada, Universitat Polit\`{e}cnica de Val\`{e}ncia, 46022 (Val\`{e}ncia), Spain}
\author{Fran\c{c}ois Leo}
\affiliation{Service OPERA-\textit{Photonique}, Universit\'e libre de Bruxelles (U.L.B.), 50~Avenue F. D. Roosevelt, CP 194/5, B-1050 Brussels, Belgium}
\author{Simon-Pierre Gorza}
\email{simon.pierre.gorza@ulb.be}
\affiliation{Service OPERA-\textit{Photonique}, Universit\'e libre de Bruxelles (U.L.B.), 50~Avenue F. D. Roosevelt, CP 194/5, B-1050 Brussels, Belgium}

\begin{abstract} 
Temporal cavity solitons (CSs) are stable, localized particle-like objects in the form of optical pulses that circulate indefinitely in coherently driven nonlinear resonators. 
In the spectral domain, they form highly coherent frequency combs. Owing to their remarkable stability, they are attracting attention for applications in sensing, metrology, or optical signal synthesis.
In this work, we report on the dynamics of CSs interacting with a trapping potential. We demonstrate that this interaction provides a powerful means to control their properties such as position, speed, and central frequency. 
Our theoretical analysis predicts fundamental limitations on the spectral shift of CSs relative to the driving frequency. 
Specifically, it reveals that within a broad range of detunings, frequency-shifted CSs encounter destabilization through a Hopf bifurcation.
Moreover, we find that with periodic potentials, the Kelly sidebands emitted by trapped solitons undergo Bloch oscillations. In our experiments, we use an intracavity phase modulator to create the equivalent of an external real potential. We observe stable blue- and red- shifted solitons up to a limit close to our theoretical predictions. We then show theoretically and experimentally that this unprecedented level of control over the CS spectrum can be leveraged to cancel the Raman-induced self-frequency shift and even to stabilize CSs beyond the limitation imposed by stimulated Raman scattering. Our results provide valuable insights for applications requiring robust and potentially rapid tunable control over the cavity soliton properties. 

\end{abstract}

\maketitle

\section{INTRODUCTION}
Solitons are waveforms that maintain their shape during propagation, owing to a delicate balance between dispersion and nonlinearity. 
First observed in hydrodynamics, they have since become a key concept in various branches of science and engineering\,\cite{allen_early_1998}, extending from plasma physics\,\cite{nakamura_observation_1984} and matter-waves\,\cite{denschlag_generating_2000} to optics\,\cite{blanco-redondo_bright_2023}.
Dissipative solitons (DSs) are a particular type of solitons encountered in non-conservative systems\,\cite{nagumo_active_1962, kapral_chemical_2012, wu_observation_1984, barland_cavity_2002}. They require an additional balance between the dissipation, inherent to the system, and gain. Besides, they represent fixed localized solutions determined by the system parameters\,\cite{grelu_dissipative_2012}. Among them, temporal cavity solitons appear in the form of optical pulses propagating endlessly in coherently driven Kerr resonators\,\cite{leo_temporal_2010, herr_temporal_2014,lilienfein_temporal_2019, wildi_dissipative_2023}. Owing to the stability of the pulse train leaving the cavity, CSs have attracted significant attention over the last decade and are used in a wide range of applications, including distance measurements\,\cite{trocha_ultrafast_2018}, high-resolution spectroscopy\,\cite{suh_microresonator_2016}, telecommunication\,\cite{lundberg_phase-coherent_2020}, and astronomy\,\cite{obrzud_microphotonic_2019}. 

Both in fundamental research and practical applications, there is value in controlling the properties of cavity solitons\,\cite{parra-rivas_effects_2014,jang_all-optical_2016,suh_microresonator_2016,erkintalo_phase_2021}.
Considerable efforts have been made in this direction, notably by introducing driving field modulations in phase\,\cite{jang_temporal_2015,jang_writing_2015,cole_kerr-microresonator_2018,weng_microresonator_2022} and amplitude\,\cite{obrzud_temporal_2017,wang_addressing_2018,nielsen_coexistence_2019,nielsen_engineered_2019}.
An alternative approach involves leveraging the interaction between nonlinear waves and an intracavity phase modulation (IPM) that plays the role of an external potential.
Interestingly, in the context of particle-like solitons, such a potential acts as an additional control parameter and leads to novel and intriguing phenomena\,\cite{strecker_formation_2002, salerno_long-living_2008, michinel_coherent_2012}. While extensively studied for conservative solitons (see e.g. \cite{moura_nonlinear_1994, serkin_nonautonomous_2007,rozenman_observation_2020} and reference therein), as well as for dissipative solitons in active mode-locked lasers\,\cite{kutz_modelocked_2006}, the dynamics of coherently driven cavity solitons in external potentials has only been recently explored, in the framework of periodic\,\cite{tusnin_nonlinear_2020,englebert_bloch_2023,he_high-speed_2023} and parabolic\cite{sun_dissipative_2022} potentials. However, the interplay between cavity solitons and trapping IPM remains, to the best of our knowledge, largely unknown\,\cite{strecker_formation_2002}.

In this work, we report on the dynamics of cavity solitons in the presence of stationary and drifting trapping potentials. We specifically demonstrate that it enables the control of the CS center frequency and the output soliton comb repetition rate.
As an application of the CS frequency control, we study the competition between an IPM and the stimulated Raman scattering (SRS). We show that the CS frequency shift induced by SRS can be fully compensated and that the trapping extends the CS existence beyond the fundamental limit imposed solely by stimulated Raman scattering\,\cite{wang_stimulated_2018}.

\section{RESULTS}

\begin{figure}
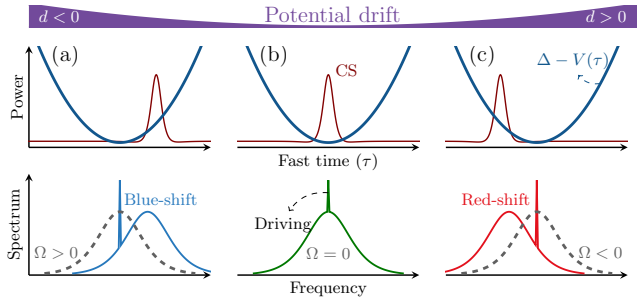
  
    \centering
    \hspace{-5mm}
    \include{Fig1_Concept}
    \caption{Trapping of a CS in a potential $V(\tau)$ for negative (a), zero (b) and positive (c) drift $d$. Without drift ($d=0$), cavity solitons are trapped at the potential, maximum corresponding to the minimum local detuning $\Delta-V(\tau)$. Stationary solitons are either blue- or red-shifted, depending on the sign of the drift.    
    }
    \label{fig:Concept}
\end{figure}

\noindent \textbf{Theoretical analysis}. We first theoretically analyze the impact of a real potential ($V$), such as that generated by intracavity phase modulation, on the existence and stability of CSs.  The starting point is the driven-dissipative nonlinear Schr\"{o}dinger equation, also known as the Lugiato-Levefer equation (LLE), a mean-field model that describes remarkably well the dynamics of coherently driven passive Kerr resonators\,\cite{lugiato_spatial_1987,haelterman_dissipative_1992,coen_universal_2013}. It is here generalized to include a potential term. As such, it can be seen as the
driven-dissipative Gross-Pitaevskii equation.
It reads in dimensionless form (see Supplemental Material\,\cite{SuppMat} for the normalization)\,\cite{rogel-salazar_grosspitaevskii_2013, sun_dissipative_2022}:

\begin{align} 
    \frac{\partial A(t,\tau)}{\partial t} = S+
    \Bigg(-1-i\left[\Delta - V(\tau)\right]+i|A|^2 \label{eq:GLLE}\\
    +\left[d\dfrac{\partial}{\partial \tau}+i\dfrac{\partial^2}{\partial\tau^2}\right] \Bigg)A,\nonumber
\end{align}
where $t$ is the (slow) time describing the evolution
of the electric field envelope $A(t,\tau)$ of a wave propagating in a dispersive resonator with anomalous group-velocity dispersion and focusing Kerr nonlinearity. $\tau$ is a (fast) time variable defined in a co-moving reference frame in which $V(\tau)$ is assumed stationary with time $t$. $S$ is the driving amplitude and $\Delta$ is the normalized phase detuning from the closest resonance. $d$ is a drift coefficient that accounts for a non-zero group velocity at the driving frequency in the co-moving reference frame.

Dissipative Kerr cavity solitons are stationary solutions that are localized along the fast time $\tau$. Provided that the potential slowly varies over the soliton envelope, it can be linearized around the soliton position $\tau=\tau_s$. Eq.\,\eqref{eq:GLLE} then reveals that $V$ modifies the local detuning experienced by the CS [i.e. $ \Delta \rightarrow \Delta-V(\tau_s)$], but also adds a linear phase $\partial_\tau V\times(\tau-\tau_s)$, hence shifting the frequency of the CS. The existence of stationary CSs in the reference frame of $V$ can therefore be understood as follows. A positive (negative) slope at the soliton location $\tau_s$ makes it travel slower (faster) than the driving frequency due to anomalous dispersion. This stabilizes the CS at the potential maximum or, equivalently, at the minimum local detuning for $d=0$. Consequently, without drift, the CS spectrum is centered on the driving frequency (see Fig.\,\ref{fig:Concept}b). 
Conversely, for $d \neq 0$, we expect the CS to settle at a position $\tau_s = \tau^*$, away from the maximum where it undergoes a frequency shift. The CS spectrum would thus be shifted towards the blue or the red, depending on the sign of the drift $d$, to achieve stationarity relative to the potential by virtue of chromatic dispersion (Figs.\,\ref{fig:Concept}a,c). This has been demonstrated in\,\cite{englebert_bloch_2023} for a very small soliton frequency shift. Yet, little is known about the existence and stability of these solitons for large drifts and therefore large frequency shifts. Numerical simulations of Eq.\ref{eq:GLLE} with different slowly varying potentials confirm that stable solitons exist up to a critical drift value.

To gain more insights into the properties of trapped CSs, we derive a reduced model by applying the Lagrangian perturbative approach to the solitary wave solutions of Eq.\,\eqref{eq:GLLE}\,\cite{matsko_timing_2013,yi_theory_2016,englebert_bloch_2023}.
Neglecting the background on which the solitons sit, we aim to describe the evolution of sech-shaped solitons whose center frequency is shifted by $\Omega$: 
$A_s(t,\tau) = B\sech\left[B(\tau-\tau_s)/\sqrt{2}\right]\exp(i\left[\phi-\Omega(\tau-\tau_s)\right])$, 
where $B$ is the soliton amplitude, $\phi$ its phase and $\tau_s$ its position along the fast time. This yields the following motion equations for the model parameters:
\begin{align}
    \frac{d\Omega}{dt} &= -\frac{\Omega}{B}\frac{dB}{dt} - 2\Omega - \dfrac{dV(\tau_s)}{d\tau_s},\label{eq:em_Omega}\\
    \frac{dB}{dt} &= -2B+\pi S\cos(\phi)\text{sech}\,\left(\frac{\Omega \pi}{\sqrt{2}B}\right),\label{eq:em_B}\\
    \frac{d\phi}{dt} &= \frac{B^2}{2}-\Omega^2-[\Delta-V(\tau_s)]-\left(\frac{d\tau_s}{dt}+d\right)\Omega,\label{eq:em_phi}\\
    \frac{d\tau_s}{dt} &= -2\Omega -d\label{eq:em_t0}.
\end{align}

We here focus on steady-state solutions. From Eq.\,\eqref{eq:em_t0}, we readily find $\Omega = -d/2$, revealing that the frequency shift is only governed by the drift. 
In dimensional units, this equation becomes $\beta_2L_c\delta\omega = \dbar t_\mathrm{R}$, where $L_c$ is the resonator length, $t_\mathrm{R}$ is the roundtrip time, $\beta_2$ is the group-velocity dispersion coefficient, $\dbar$ is the dimensional drift coefficient and $\delta\omega$ is the soliton frequency shift. It shows, as expected, that the CS is synchronized to the potential.
Defining locally the gradient of $V$ as a monotonic function $\mathcal{D} = dV(\tau_s)/d\tau_s$, the CS steady-state temporal position $\tau^*$ is found from the reciprocal function ($\mathcal{D}^{-1}$). Indeed, from Eqs.\,\eqref{eq:em_Omega} and \eqref{eq:em_t0} we have $\tau^* = \left.\mathcal{D}^{-1}\right|_{d}$.
Moreover, we see from Eq.\,\eqref{eq:em_Omega} that the frequency shift is $\Omega = -\frac{1}{2}\mathcal{D}$. Recalling that the potential is an intracavity phase modulation $\phi_\mathrm{int}(\tau)$, this frequency shift is in dimensional units $\delta\omega = -\mathcal{F}/(2\pi)\times \phi_\mathrm{int}'$  where  $\mathcal{F}$ is the cavity finesse and $'$ stands for the first derivative with respect to the fast time $\tau$. This result shows that the CS frequency shift is enhanced by the factor $\mathcal{F}/2\pi$ for intracavity modulation compared with external phase modulations of the driving, \,\cite{jang_temporal_2015} (see also Supplemental Material\,\cite{SuppMat}). This is of prime interest for applications given the recent developments of on-chip high-finesse lithium niobate microresonators that incorporate a high-speed electro-optic modulator\,\cite{he_high-speed_2023}.
Finally, we see that the frequency shift is bounded by the maximum value of the slope $|\Omega_{\SL}| = \frac{1}{2}|\mathcal{D}_{max}|$. 

The combination of Eqs.\,\eqref{eq:em_phi}-\eqref{eq:em_t0} yields the stationary soliton amplitude
\begin{equation}
B = \sqrt{2[\Delta - V(\tau^*) - \Omega^2]} = \sqrt{2(\Delta^*-\Omega^2)},
\label{eq:B}
\end{equation}
where we have introduced the local detuning $\Delta^* = \Delta - V(\tau^*)$. For $V=0$, we retrieve the usual CS amplitude ($B=\sqrt{2\Delta}$)\,\cite{coen_universal_2013}. For static potentials ($d=0\Rightarrow\Omega = 0$), Eq.\,\eqref{eq:em_Omega} shows that CSs are located at the extrema of $V$. However, only the maxima are stable.
Consequently, CSs are trapped at minimum local detunings $\Delta^* = \Delta - \max[V(\tau^*)]$, which sets the soliton amplitude accordingly. Eventually, for drifting potentials ($d\neq 0\Rightarrow\Omega = -d/2$), the amplitude is further reduced because the actual detuning of the soliton is shifted from the detuning at the driving frequency as a result of the frequency shift and the dispersion. The CS amplitude and width are thus given by the effective detuning $\Delta_\mathrm{eff} = \Delta^*-\Omega^2$.

Finally, the limit of existence of CSs, is found by setting $\cos(\phi)=1$ in Eq.\,\eqref{eq:em_B}: 
\begin{equation}
    \frac{2}{\pi} \sqrt{2(\Delta_{\DL}^*-\Omega_{\DL}^2)} = S\,\text{sech}\left(\dfrac{\pi\Omega_{\DL}}{2\sqrt{\Delta_{\DL}^*-\Omega_{\DL}^2}}\right).
    \label{eq:SNs}
\end{equation}
The maximum detuning ($\Delta_{\DL}^*$) can be seen as the pump depletion limit and corresponds to a saddle-node bifurcation. The predicted value by the LLE [$\Delta_{\DL}^{\text{LLE}}=(\pi\,S)^2/8$] is once more retrieved for $V = 0$. This result is generalized for static potentials by considering the local detuning $\Delta^*$. 
In addition, this limit is lowered for $d \neq 0$ because of the frequency shift. 
Rewriting Eq.\,\eqref{eq:SNs} with the effective detuning $\Delta_\mathrm{eff}$, the right-hand side of this equation suggests that the driving strength $S$ is effectively diminished by the normalized spectral amplitude of the soliton at the driving frequency.

\begin{figure} 
    \centering
    \hspace{-5mm}\includegraphics[width=\linewidth]{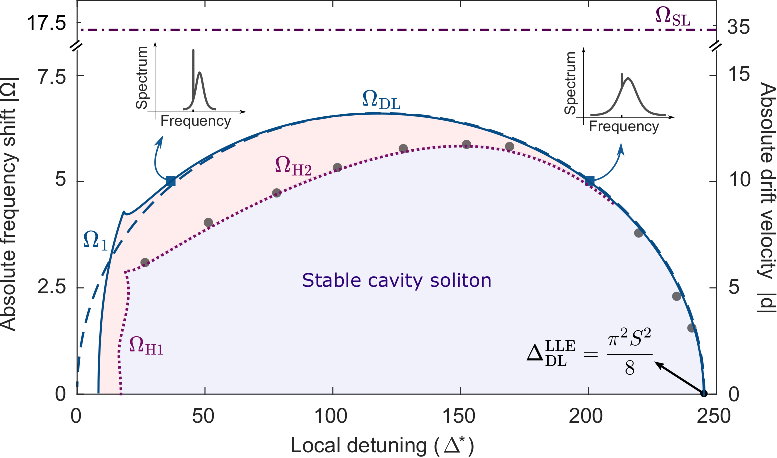}
    \caption{Phase diagram in the ($\Delta^*,|\Omega|$)-parameter space showing the region of existence of cavity solitons and their stability for a cosine potential 
   ($J=67.7$, $\mathcal{W} = 0.52$), and $S=14.1$. The horizontal dash-dotted line ($\Omega_{SL}$) is the limit set by the maximum slope of $V$. The dashed blue line is the driving depletion limit of CS ($\Omega_\DL$) given by the reduced model Eq.\,\eqref{eq:SNs}, while the two plain blue lines are the saddle-node bifurcations ($\Omega_1$ and $\Omega_\DL$) of CS solutions of Eq.\,\eqref{eq:GLLE}. 
  The Hopf bifurcations ($\Omega_\mathrm{H 1,2}$) at which CSs gain or lose their stability are represented by the purple dotted lines. The gray dots show the frequency shift where the CS ceases to exist in direct numerical simulations of Eq.\,\eqref{eq:GLLE} when slowly increasing the drift $d$. The blue and red shaded areas indicate regions of stable and unstable CSs, respectively. All numerical continuations were performed with the Julia package BifucationKit\,\cite{veltz_bifurcationkitjl_2020}.}
    \label{fig:Freq_Shift}
\end{figure}

We now particularize to a periodic potential of the form $V(\tau) = J\cos(\mathcal{W}\tau)$, where we set the amplitude and frequency to $J=67.7$ and $\mathcal{W}\approx 0.52$, respectively, for comparison with our experiments. The existence region of cavity solitons, plotted in the two-dimensional ($\Delta^*,|\Omega|$)-parameter space, is reported in Fig.\ref{fig:Freq_Shift}.
The limitation from the maximum slope is $|d_\SL|=2|\Omega_{\SL}|$. 
However, with our parameters, it cannot be reached because the depletion limit is met first. Furthermore, as seen in the figure, the largest frequency shift, determined by the depletion limit ($|\Omega_\DL|$), shows a maximum. On the one hand, at large detunings, CSs are spectrally broad but close to the depletion limit without drift (i.e. for $\Omega=0$). A small decrease of the effective driving strength due to the frequency shift is thus sufficient to cross their existence limit. On the other hand, in the small detuning range, the spectral width of CSs is initially small. The effective driving strength is therefore quickly reduced by the frequency shift, all the more so as the spectral width is also narrowed by the shift. 

To confirm the depletion limit from Eq.\,\eqref{eq:SNs}, we also compute this limit by finding the corresponding saddle-node solution of Eq.\,\eqref{eq:GLLE} using a continuation scheme. The agreement is excellent except for low detunings where the homogenous background of the CS solution is not negligible. 

We then look at the stability of the localized solutions. Interestingly, we find that CSs can undergo a Hopf bifurcation below the depletion limit. This bifurcation (denoted as $\Omega_{\mathrm{H 2}}$) is not found with our reduced model from the motion equations, probably because the chirp is not included in the ansatz\,\cite{longhi_variational_1997}. This new bifurcation connects to the well-known Hopf bifurcation ($\Omega_{\mathrm{H 1}}$) that stabilizes the CS solutions as the detuning increases.     
Simulations of Eq.\,\eqref{eq:GLLE} reveal that CSs cease to exist abruptly upon crossing the frequency shift threshold $|\Omega_{\mathrm{H 2}}|$. Hence, there is a maximum frequency shift CSs can withstand before they disappear. It is set either by the depletion limit or the Hopf bifurcation. 
We note that for much smaller maximum slope of $V$, the $|\Omega_{\SL}|$ threshold could be reached first as in \,\cite{englebert_bloch_2023}. Beyond $\Omega_{\SL}$, the existence limits given by $\Omega_{\DL}$ and $\Omega_{\mathrm{H 2}}$ are thus not relevant, as they only concern steady-state CSs.

\begin{figure} 
    \centering
    \hspace{-5mm}\includegraphics[width=\linewidth]{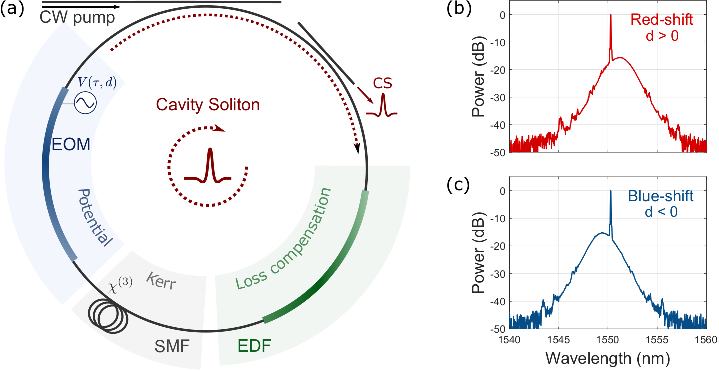}
    \caption{(a) Schematic of the experimental set-up. The coherently driven resonator is made of single-mode fiber (SMF) and includes an intracavity electro-optic phase modulator (EOM) to generate the periodic potential. The erbium-doped fiber (EDF) amplifier partially compensates for the cavity loss. Typical spectra measured for positive (b) or negative (c) potential drifts. The drift amplitude and sign control the CS spectral position, which can be blue or red-shifted with respect to the pump.}     
    \label{fig:setup}
\end{figure}


\vspace{12pt}
\noindent \textbf{Experimental setup.} We aim next to experimentally confirm our theoretical predictions. A schematic of the setup is shown in Fig.\,\ref{fig:setup} (see also Supplemental Material\,\cite{SuppMat}). It mainly consists of a $L_c=64$\,m-long single-mode fiber ring resonator (3.12\,MHz free-spectral range), incorporating an electro-optic modulator (EOM) to generate a periodic potential. It is driven with a radio-frequency harmonic signal whose amplitude and frequency can be tuned. The resonator also includes a short piece of erbium-doped fiber to partially compensate the cavity loss\,\cite{englebert_temporal_2021}. The resulting effective loss is $\Lambda_e=3\%$ (effective finesse $\mathcal{F}_e = 206$).
The resonator is driven by a highly coherent continuous-wave (CW) laser at 1550.12\,nm. 
We note that despite the high effective finesse and CW driving, there is no isolator in the resonator to prevent Brillouin lasing because the intracavity phase modulation raises the Brillouin laser threshold above the continuous background powers on which CSs sit. 
Finally, a counter-propagating small signal, which can be frequency-shifted, is used to phase stabilize the detuning between the cavity and the driving laser.

\begin{figure} 
    \centering
    \hspace{-5mm}\includegraphics[width=\linewidth]{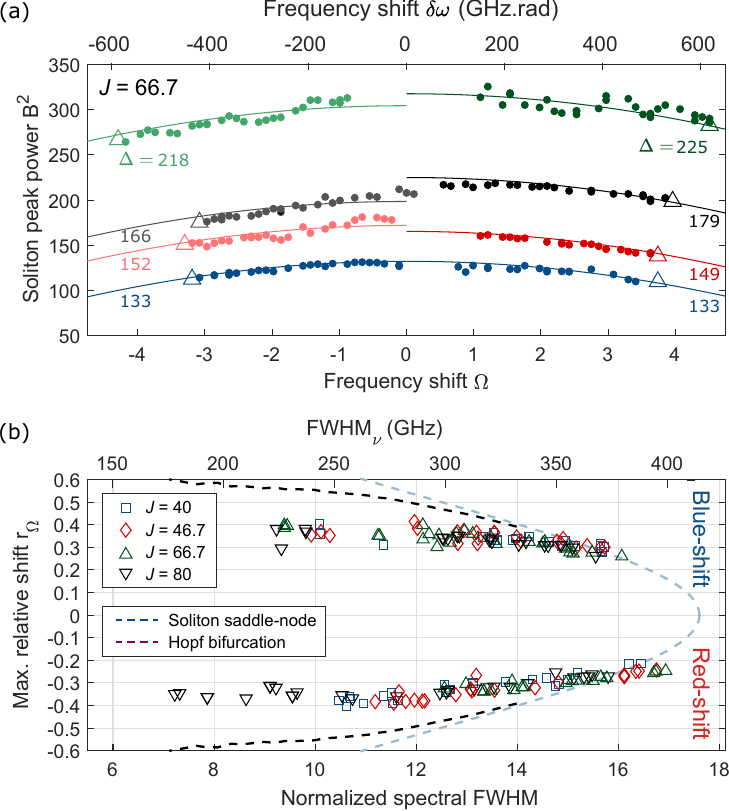}
    \caption{(a) Experimental soliton peak power (circle) as a function of the frequency shift $\Omega$ for $S=14.1$. Plain lines: analytical solutions [see Eq.\,\eqref{eq:B}]. 
    The number below the curves corresponds to the normalized detuning $\Delta$. 
    The last measurement before the soliton disappears is indicated by a triangle and is reported in panel (b). 
   (b) Maximum relative shift $r_\Omega = \Omega/(2\pi \times \mathrm{FWHM}_\nu)$ as a function of the soliton spectral full width at half maximum (FWHM$_\nu$) for different modulation amplitude $J$. 
    To delimit the theoretical soliton region of existence, the depletion limit and the Hopf bifurcation are also shown (see Fig.\,\ref{fig:Freq_Shift}).}
    \label{fig:Freq_Shift_Exp}
\end{figure}

\noindent \textbf{Temporal trapping by drifting potentials.} In a first experiment, we set the driving to $S=14.1$ and the amplitude of the IPM to $J=66.7$ to investigate the soliton dynamics as the drift coefficient is linearly scanned. We start with a frequency modulation ($\mathcal{W} = 0.52$) that matches an exact integer multiple of the cavity FSR to generate a stationary phase modulation ($d=0$, see Supplemental Material\,\cite{SuppMat}). Once the cavity is stabilized at a given phase detuning, we excite a single CS using a single addressing pulse\,\cite{leo_temporal_2010,englebert_high_2023}. After the transient, we slowly tune the frequency of the modulation to higher ($d<0$) or lower ($d>0$) values until the CS disappears. For each modulation frequency, we record the CS spectrum (see examples in Fig.\,\ref{fig:setup} as well as a full scan in the Supplemental Material\,\cite{SuppMat}) from which we measure the CS spectral width ($\propto B$) and frequency shift ($\Omega$). Fig.\,\ref{fig:Freq_Shift_Exp}(a) shows the evolution of $B^2$ as a function of the frequency shift, for different values of the detuning and for both positive and negative drifts. The results match well with the analytical predictions from Eq.\,\eqref{eq:B} with $\tau^* = \left.\mathcal{D}^{-1}\right|_{-2\Omega}$ and confirm the reduction of the peak power ($\propto B^2$) and spectral width of CSs with the frequency shift $|\Omega|$, i.e. with the drift.
The results shown in Fig.\,\ref{fig:Freq_Shift_Exp}(b) represent the measured maximum frequency shift, normalized to the CS spectral width. They are plotted as a function of the CS spectral width and compared with the depletion limit ($|\Omega_\DL|$, Eq.\ref{eq:SNs}) and the Hopf bifurcation ($\Omega_\mathrm{H 2}$). We confirm the existence of a limit frequency shift and thus a maximum drift of the potential the CS can withstand. The measured values of the limit relative spectral shift are all in the range [0.3-0.4] (see Supplemental Figure S5\,\cite{SuppMat} for experimental data acquired at different driving powers). They are also independent of the periodic modulation amplitude $J$, as expected for the amplitudes considered. The comparison with the theoretical predictions shows, however, that in the experiments the CSs does not always reach the theoretical maximum spectral shift. This discrepancy can likely be explained by the reduction of the robustness of CSs against perturbations as they approach their existence limit.    
Our results confirm the impact of the intracavity phase modulation on the existence and characteristics of trapped CSs. Particularly, the modulation enables to control the CS central wavelength as well as the repetition rate of the frequency comb leaving the cavity, which may open up interesting new avenues\,\cite{he_high-speed_2023}.  
In what follows, as an illustration, we show how such control can be harnessed to overcome the detrimental effects of Raman scattering.

\begin{figure} 
    \centering
    \hspace{-5mm}\includegraphics[width=\linewidth]{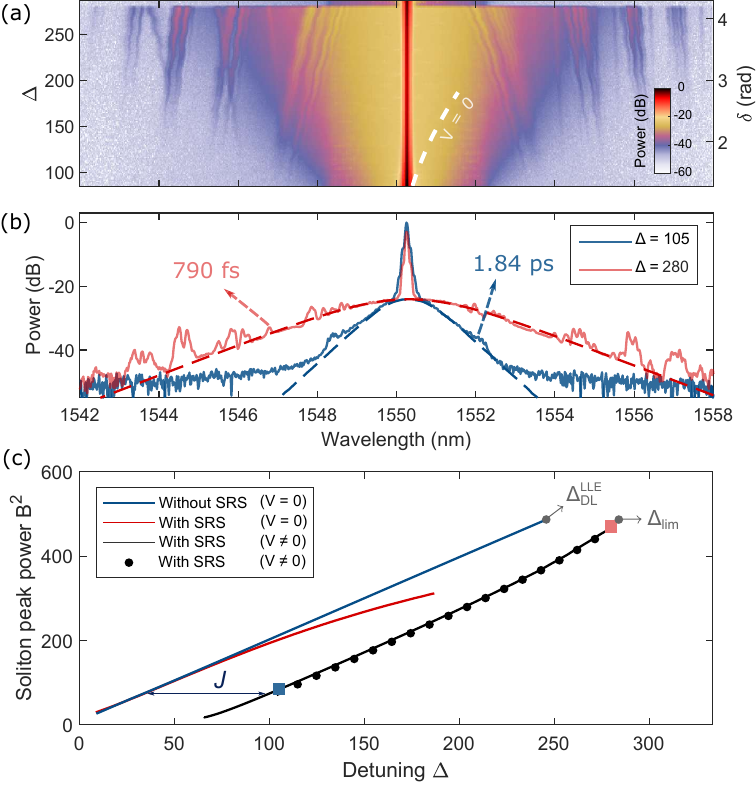}
    \caption{ (a) Two-dimensional map of experimentally recorded spectra of the solitons trapped in a stationary intra-cavity phase modulation as a function of the detuning. The white dashed line shows the theoretically predicted CS central wavelength and range of existence without modulation ($V=0)$. (b) Two experimental spectra corresponding to different cavity detunings $\Delta$ (solid lines), and predicted CS spectra by numerical integration of the mean-field model with $S=14.1$, $J=66.67$, $\mathcal{W}=0.52$ and $\tau_R=4.3\times10^{-4}$ (dashed lines). (c) Intracavity peak power of stable CSs ($|B|^2$) as a function of the detuning. The black curve represents the solutions of the mean-field model with SRS (solid) and the 
    circles indicate the result of the motion equations. The red and blue curves show the solutions for $V=0$ with and without SRS, respectively. We note that the black curves are shifted by the modulation amplitude $J$ for small detunings, i.e., small soliton bandwidths, as the CS remains located close to the minimum local detuning $\Delta^*\approx \Delta -J$. The squares are the experimental values reported in (b).       
    }
    \label{fig:Raman_Cancel}
\end{figure}


\vspace{12pt}
\noindent \textbf{Competition between temporal trapping and Raman scattering.} In resonators made of amorphous materials, stimulated Raman scattering (SRS) is a significant effect that red-shifts the spectrum of cavity solitons relative to the driving laser wavelength
and whose magnitude depends on the detuning\,\cite{karpov_raman_2016, yi_theory_2016}. This frequency shift, in turn, changes the soliton repetition rate, a coupling mechanism that is beneficial to filter out the frequency noise of the driving laser, though only in a limited spectral bandwidth\cite{lei_optical_2022}, otherwise, it is detrimental. In addition, SRS is known to set an upper limit to the range of detunings over which CSs exist, owing to the onset of a Hopf bifurcation. This consequently restricts their duration and bandwidth.\,\cite{wang_stimulated_2018}. However, assuming that the SRS limitation finds its origin in the frequency shift, it should be possible to overcome this limit by trapping CSs within potentials. Indeed, the synchronization of a steady-state CS to an intracavity phase modulation without drift
compels its spectrum to remain centered on the driving. To experimentally confirm this prediction, we use the same driving power and potential shape as in the previous experiments, but we now scan the detuning $\Delta$ 
with $d=0$. Our results are shown in Fig.\,\ref{fig:Raman_Cancel}(a)-(b). They confirm that stationary potentials do not prevent the existence of stable CSs in the presence of strong SRS. 
Moreover, with increasing detuning, the center of the CS spectrum remains fixed at the driving wavelength, even as the soliton duration decreases and its spectrum widens.   

In order to have a better understanding of the competition between SRS and an intracavity phase modulation, we first extend our previous Lagrangian analysis to include the Raman scattering. To this end, we add the term $-i\tau_R A \partial_\tau|A|^2$ in the right-hand side of Eq.\,\eqref{eq:GLLE}, 
with $\tau_R$ the normalized Raman time constant (see Supplemental Material\,\cite{SuppMat}). 
We note that we here consider the linear approximation of the Raman spectral gain, which is valid given the typical duration of CSs in fiber resonators ($>100$\,fs)\,\cite{atieh_measuring_1999}. In the steady-state, the motion equation for the soliton position $\tau_s$ still gives $\Omega = -d/2$. For stationary IPM, we thus have $\Omega =0$, and the motion equation for the frequency shift gives:   
\begin{equation}
    - \dfrac{dV(\tau_s)}{d\tau_s}|_{\tau^*}=J\mathcal{W}\sin(\mathcal{W}\tau^*) = 4\tau_RB^4/15.
    \label{eq:RamanReducedEqs}
\end{equation}
It shows that the CS stabilizes at a position $\tau^*$ such that the corresponding potential-induced spectral blue-shift exactly compensates the Raman soliton red-shift (see also Supplemental Material\,\cite{SuppMat} for the evolution of the CS spectrum over one roundtrip). Finally, the CS amplitude ($B$) is found by combining this latter equation with the Eq.\,\eqref{eq:B}, which is unmodified by the SRS. The solutions from the motion equations are shown in Fig.\,\ref{fig:Raman_Cancel}(c). An excellent agreement with the numerical continuation of the solutions of the mean-field equation is obtained.
The comparison with and without IPM reveals that higher peak powers and thus shorter CSs can be obtained with SRS when they are trapped in the modulation.
In the experiment, we measure a 790\,fs CS persisting endlessly in the resonator, while stable CSs are limited to 988\,fs by the SRS without the potential. 
(We note that we have not performed measurements at $V=0$ because of the Brillouin lasing.)     
Yet, with the experimental parameters, the limit detuning ($\Delta_\mathrm{lim}$) is here set by the depletion limit (equation \ref{eq:SNs} for $\Omega_\mathrm{DL}=0$). The shortest CS is thus identical to the one given by the LLE.    

\begin{figure} 
    \centering
    \vspace{2mm}
   \includegraphics[width=\linewidth]{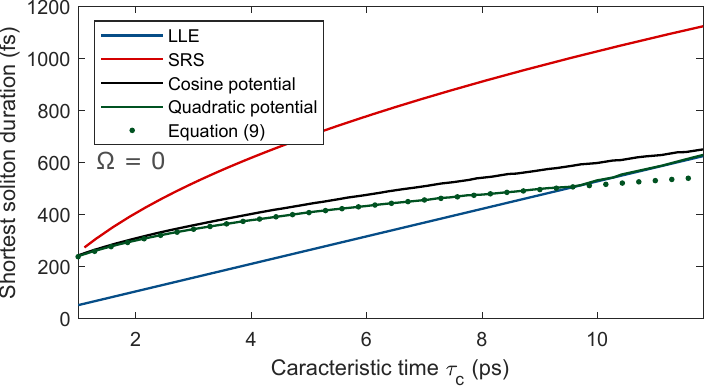}
    \caption{Shortest cavity soliton duration as a function of the characteristic times $\tau_c$ with $S=30$. Solid curves correspond to the results from numerical continuations of solutions of the mean-field model. The dots show the limit given by Eq.\,\ref{eq:limitQuadPotential}. The parameters for the potentials are $J=209$, $\omega_\mathrm{RF}=2\pi\times20$\,GHz and $d=0$. The Raman time constant is the value typical for silica fibers, $\tauRbar=3$\,fs. We note that $\tau_c$=7\,ps in the experiments.
    }
    \label{fig:Raman_Limit}
\end{figure}

Equation \ref{eq:RamanReducedEqs} suggests that the LLE limit could always be reached providing that the potential is sufficiently steep.  To broaden our analysis, we investigate how far the balance between the Raman red-shift and the potential-induced blue-shift can be maintained. We find that at large detunings, CSs can undergo a new saddle-node bifurcation before reaching the depletion limit. 
Figure \ref{fig:Raman_Limit} shows for $S=30$ the duration of the shortest stable CS as a function of the cavity characteristic time, $\tau_c = \sqrt{-\beta_2L_c/\Lambda_e}$, for a large amplitude harmonic IPM, as well as for its parabolic approximation $V(\tau) = J[1-(\mathcal{W}\tau)^2/2]$. 
These results are compared to the corresponding shortest CS without trapping IPM both in the absence and presence of SRS. We see that the modulation enables the CS to approach or even reach the standard limit without SRS. Yet, as $\tau_c$ decreases, the compensation of the self-frequency shift becomes less effective. Ultimately, the confinement in the IPM does not suffice to stabilize the CS beyond the Raman limit.
Interestingly, for stationary quadratic modulations, the motion equations can be solved to find this new limit (See Supplemental Material\,\,\cite{SuppMat}). In dimensional units, it reads:
\begin{equation}
    \Delta\tau_\mathrm{FWHM, min} \approx 2.02\left( \frac{\alpha_e\tauRbar^2\tau_c^2}{J_\mathrm{RF} \omega_\mathrm{RF}^2}\right)^{1/6},
    \label{eq:limitQuadPotential}
\end{equation}
where $\tauRbar = \tau_r\tau_c$ is the Raman time constant, $\omega_\mathrm{RF} = \mathcal{W}/\tau_c$ is the modulation frequency and $J_\mathrm{RF} = J \Lambda_e/2$ its amplitude. An excellent agreement with the numerical solutions of the mean-field model is obtained for $\tau_c$ above $\approx 2$\,ps.

\begin{figure}
    \centering
    \vspace{2mm}
   \includegraphics[width=\linewidth]{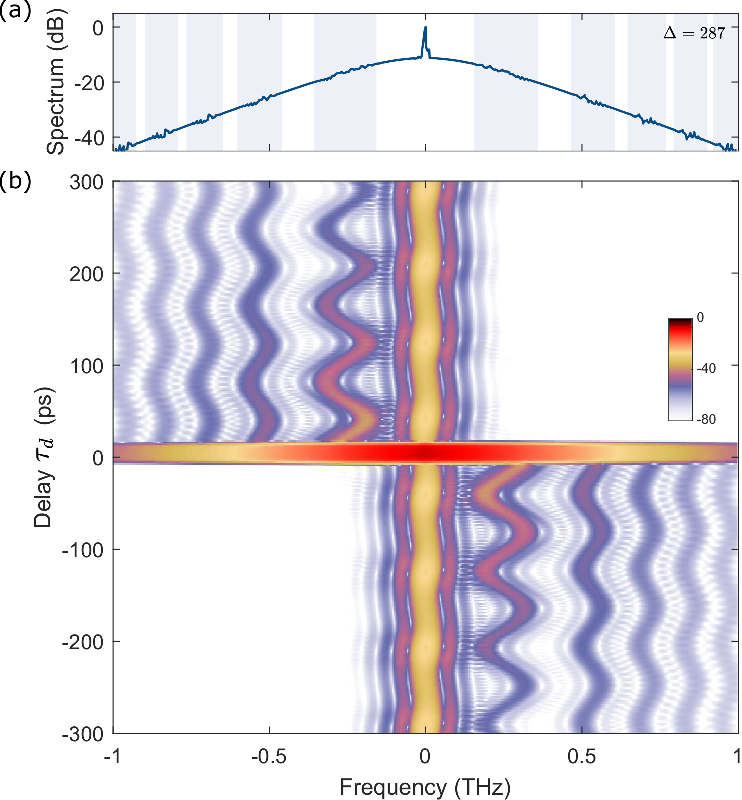}
    \caption{(a) Spectrum of a CS displaying structured Kelly sidebands (50\,pm resolution) due to periodic intracavity modulation. The shadowed regions highlight the extension of the Kelly sidebands. (b) Corresponding spectrogram of the intracavity field (gate function: 20\,ps-wide super-Gaussian of order 6). The vertical line at the center is the background, while the soliton spectrum is seen as the horizontal band around the zero delay. The parameters are identical to the experiment reported in Fig.\ref{fig:Raman_Cancel}(b) for $\Delta=280$.      
    } 
    \label{fig:Spectro}
\end{figure}

\vspace{12pt}
\noindent \textbf{Bloch oscillations of the Kelly sidebands.} Perturbations over the roundtrip of a CS and its subsequent reshaping into a fundamental soliton result in the emission of linear waves\cite{luo_resonant_2015}. Typically, these are observed as narrowband spectral features, known as Kelly sidebands, with their location satisfying the phase-matching relation $[\beta(\omega)-\beta(\omega_0)-\beta_1\times(\omega-\omega_0)]L_c = 2\pi m +\delta_0$, where $\beta$ is the propagation constant, $\beta_1$ is the inverse of the group velocity, $\omega_0$ is the CS central frequency, $\delta_0$ is the cavity detuning in radian and $m$ is an integer. In stark contrast, the spectra reported in Fig.\ref{fig:Raman_Cancel} display broad and highly structured features in the CS spectral wings. Figure \ref{fig:Spectro}(a) shows the spectrum of a CS numerically simulated with a lumped-element model of our cavity. It confirms that ultra-stable CSs display structured Kelly sidebands in the presence of periodic potentials. Yet, their width and central location cannot be accurately predicted by a generalized phase-matching condition that accounts for the modulation in the fast time: $\delta_0 \rightarrow \delta_0-\phi_\mathrm{int}(\tau)$ in the previous equation, where $\phi_\mathrm{int}(\tau)$ is the intracavity phase modulation in physical units (see Supplemental Material\,\cite{SuppMat}). To gain more insight, Fig.\ref{fig:Spectro}(b) shows the spectrogram of the intracavity field. It reveals that the complex shape of the Kelly-bands in the spectrum comes from the interference of very regular oscillations of the Kelly waves in the time-frequency domain. These oscillations, which originate from the relative drift between the linear waves and the intracavity phase modulation, can be seen as Bloch oscillations (BOs) along a synthetic frequency dimension\cite{chen_real-time_2021}. However, contrary to BO experiments with linear wave packets in recirculation loops\cite{chen_real-time_2021} or CSs in driven resonators\cite{englebert_bloch_2023}, the linear (Kelly) waves are here continuously generated by the soliton. Hence, this results in a stationary pattern, round-trip after round-trip. The effective force responsible for the oscillatory motion in the synthetic frequency lattice is $F = \Delta \nu / (n\times \mathrm{FSR})$, where 
$\Delta \nu = n\times \mathrm{FSR}-\nu_\mathrm{RF}$
with $\nu_\mathrm{RF}$ the frequency of the phase modulation and $n$ the closest integer multiple between $\nu_\mathrm{RF}$ and the FSR\,\cite{englebert_bloch_2023}. Yet, owing to the dispersion, for a stationary phase modulation at $\omega_0$ ($d=0$), $F$ is directly proportional to $\Delta\omega = \omega-\omega_0$, where $\omega$ is the frequency of the sideband. Since the amplitude of BOs is given by $A_\mathrm{BO} = J_\mathrm{RF} / F$, this explains the decrease in the amplitude of the spectral oscillations with the distance of the spectral sideband from the central CS position seen in Fig.\ref{fig:Spectro}(b). More precisely, the oscillation amplitudes from the simulation agree very well with the $1/\Delta\omega$ scaling expected for BOs (see Supplemental Material\,\,\cite{SuppMat}).         

\section{DISCUSSION}

In this work, we have studied how the unique particle-like characteristics of solitons and their interaction with slowly varying trapping potentials can be leveraged to control the properties of cavity solitons. 
In our experiments, the potential is generated by intracavity phase modulation at a frequency close to a high harmonic of the free spectral range of our fiber resonator. We have shown that the locking of the cavity soliton to the modulation enables to up- or dow-shift its spectrum with respect to the driving frequency by desynchronizing the modulation to the cavity round-trip time. 
We theoretically found that for deep potentials, the limit spectral shift ($\delta\omega_\mathrm{max}$) is fundamentally set either by the driving pump depletion limit or a Hopf bifurcation, while it is set by the maximum potential slope for shallow potentials\,\cite{englebert_bloch_2023}. For temporal cavity solitons, the maximum shift, in turn, sets a limitation on the tuning of the output soliton comb repetition rate: $\Delta \mathrm{FSR}_\mathrm{max} = -\beta_2 L_c \mathrm{FSR}^2 \delta\omega_\mathrm{max}$. 

Experimentally, we have investigated the trapping by intracavity phase modulations. We have observed that the soliton spectrum can robustly be spectrally blue- or red-shifted and that the reduction of the CS amplitude with the frequency shift is very well predicted by our theoretical model. The experimental maximum relative frequency shifts globally agree with the theoretical predictions but seem restricted to about 0.4, which could indicate a fundamental practical limitation.

We then demonstrated that the ability to deterministically manipulate the soliton frequency can be exploited to overcome the soliton self-frequency shift arising from the stimulated Raman scattering (SRS). We observed the formation of potential-stabilized solitons that would not otherwise have existed\,\cite{wang_stimulated_2018}. However, we theoretically showed that there is a minimum soliton duration below which the balance between the potential induced blue-shift and the Raman red-shift cannot be maintained.   

Previous works have shown that cavity solitons can also be trapped and manipulated by external phase modulations \cite{jang_temporal_2015, cole_kerr-microresonator_2018} of the coherent driving. Yet, compared to external modulations, intracavity phase modulations enhance the tuning range of the CS spectral shift, and thus of the soliton comb repetition rate, by the inverse of the cavity loss $\Lambda_e$ (see Supplemental Material\,\cite{SuppMat}). The improvement can be significant for low-loss resonators, emphasizing the importance of dissipation on the dynamics of CSs in internally modulated resonators.  

We envision intracavity phase modulation to play an important role in the future of cavity soliton based frequency combs. On the one hand, the robust locking mechanism can be leveraged for generating ultralow-noise optical combs and microwave signals. This can be achieved through direct synchronization of the cavity solitons to an external microwave reference\cite{he_high-speed_2023} or by employing a coupled optoelectronic oscillator scheme\,\cite{pedaci_positioning_2006,yao_dual_1997,yu_ultralow-noise_2005}.
On the other hand, the motion equations show that the potential gradient governs the rate at which the soliton frequency can be changed per round trip. The ability to rapidly vary the soliton frequency and, thus, the soliton comb repetition rate by steep potentials can benefit many applications, such as LiDAR and sensing.   
%
Finally, our findings could shed light on recently reported results in internally modulated cavity soliton microcombs \,\cite{he_high-speed_2023}, and could be extended to solitary waves of the parametrically driven nonlinear Schr\"{o}dinger equation in optics \cite{englebert_parametrically_2021, moille_parametrically_2024} and hydrodynamics\cite{wu_observation_1984}, for which the potential will interact with the attraction or repulsion force in soliton pairs\,\cite{rozenman_amplitude_2019}. 
\section*{ACKNOWLEDGMENTS}
This work was supported by the “Fonds de la Recherche Scientifique” (F.R.S.-FNRS) and the FWO under the Excellence of Science (EOS, 40007560) program, by the F.R.S.-FNRS (CDR, J.0079.23) and by the European Research Council (ERC) under the Horizon Europe research and innovation program (Grant Agreement No. 101113343). N.E. acknowledges support from the Belgian American Educational Foundation (B.A.E.F.). N.E., C.S., C.M.A. and F.L. acknowledge the support of the F.R.S.-FNRS.

\bibliography{Ref} 

\begin{thebibliography}{10}

\bibitem{allen_early_1998}
J.~E. Allen.
\newblock The {Early} {History} of {Solitons} ({Solitary} {Waves}).
\newblock {\em Physica Scripta}, 57(3):436, March 1998.

\bibitem{nakamura_observation_1984}
Y.~Nakamura and I.~Tsukabayashi.
\newblock Observation of {Modified} {Korteweg}---de {Vries} {Solitons} in a
  {Multicomponent} {Plasma} with {Negative} {Ions}.
\newblock {\em Physical Review Letters}, 52(26):2356--2359, June 1984.

\bibitem{denschlag_generating_2000}
J.~Denschlag, J.~E. Simsarian, D.~L. Feder, Charles~W. Clark, L.~A. Collins,
  J.~Cubizolles, L.~Deng, E.~W. Hagley, K.~Helmerson, W.~P. Reinhardt, S.~L.
  Rolston, B.~I. Schneider, and W.~D. Phillips.
\newblock Generating {Solitons} by {Phase} {Engineering} of a {Bose}-{Einstein}
  {Condensate}.
\newblock {\em Science}, 287(5450):97--101, January 2000.

\bibitem{blanco-redondo_bright_2023}
Andrea Blanco-Redondo, C.~Martijn de~Sterke, Chris Xu, Stefan Wabnitz, and
  Sergei~K. Turitsyn.
\newblock The bright prospects of optical solitons after 50 years.
\newblock {\em Nature Photonics}, 17(11):937--942, November 2023.

\bibitem{nagumo_active_1962}
J.~Nagumo, S.~Arimoto, and S.~Yoshizawa.
\newblock An {Active} {Pulse} {Transmission} {Line} {Simulating} {Nerve}
  {Axon}.
\newblock {\em Proceedings of the IRE}, 50(10):2061--2070, October 1962.

\bibitem{kapral_chemical_2012}
Raymond Kapral and K.~Showalter.
\newblock {\em Chemical {Waves} and {Patterns}}.
\newblock Springer Science \& Business Media, December 2012.

\bibitem{wu_observation_1984}
Junru Wu, Robert Keolian, and Isadore Rudnick.
\newblock Observation of a {Nonpropagating} {Hydrodynamic} {Soliton}.
\newblock {\em Physical Review Letters}, 52(16):1421--1424, April 1984.

\bibitem{barland_cavity_2002}
Stephane Barland, Jorge~R. Tredicce, Massimo Brambilla, Luigi~A. Lugiato,
  Salvador Balle, Massimo Giudici, Tommaso Maggipinto, Lorenzo Spinelli,
  Giovanna Tissoni, Thomas Knödl, Michael Miller, and Roland Jäger.
\newblock Cavity solitons as pixels in semiconductor microcavities.
\newblock {\em Nature}, 419(6908):699--702, October 2002.

\bibitem{grelu_dissipative_2012}
Philippe Grelu and Nail Akhmediev.
\newblock Dissipative solitons for mode-locked lasers.
\newblock {\em Nature Photonics}, 6(2):84--92, February 2012.

\bibitem{leo_temporal_2010}
François Leo, Stéphane Coen, Pascal Kockaert, Simon-Pierre Gorza, Philippe
  Emplit, and Marc Haelterman.
\newblock Temporal cavity solitons in one-dimensional {Kerr} media as bits in
  an all-optical buffer.
\newblock {\em Nature Photonics}, 4(7):471--476, July 2010.

\bibitem{herr_temporal_2014}
T.~Herr, V.~Brasch, J.~D. Jost, C.~Y. Wang, N.~M. Kondratiev, M.~L. Gorodetsky,
  and T.~J. Kippenberg.
\newblock Temporal solitons in optical microresonators.
\newblock {\em Nature Photonics}, 8(2):145--152, February 2014.

\bibitem{lilienfein_temporal_2019}
N.~Lilienfein, C.~Hofer, M.~Högner, T.~Saule, M.~Trubetskov, V.~Pervak,
  E.~Fill, C.~Riek, A.~Leitenstorfer, J.~Limpert, F.~Krausz, and I.~Pupeza.
\newblock Temporal solitons in free-space femtosecond enhancement cavities.
\newblock {\em Nature Photonics}, 13(3):214--218, March 2019.

\bibitem{wildi_dissipative_2023}
Thibault Wildi, Mahmoud~A. Gaafar, Thibault Voumard, Markus Ludwig, and Tobias
  Herr.
\newblock Dissipative {Kerr} solitons in integrated {Fabry}–{Perot}
  microresonators.
\newblock {\em Optica}, 10(6):650--656, June 2023.

\bibitem{trocha_ultrafast_2018}
P.~Trocha, M.~Karpov, D.~Ganin, M.~H.~P. Pfeiffer, A.~Kordts, S.~Wolf,
  J.~Krockenberger, P.~Marin-Palomo, C.~Weimann, S.~Randel, W.~Freude, T.~J.
  Kippenberg, and C.~Koos.
\newblock Ultrafast optical ranging using microresonator soliton frequency
  combs.
\newblock {\em Science}, 359(6378):887--891, February 2018.

\bibitem{suh_microresonator_2016}
Myoung-Gyun Suh, Qi-Fan Yang, Ki~Youl Yang, Xu~Yi, and Kerry~J. Vahala.
\newblock Microresonator soliton dual-comb spectroscopy.
\newblock {\em Science}, 354(6312):600--603, November 2016.

\bibitem{lundberg_phase-coherent_2020}
Lars Lundberg, Mikael Mazur, Ali Mirani, Benjamin Foo, Jochen Schröder, Victor
  Torres-Company, Magnus Karlsson, and Peter~A. Andrekson.
\newblock Phase-coherent lightwave communications with frequency combs.
\newblock {\em Nature Communications}, 11(1):201, January 2020.

\bibitem{obrzud_microphotonic_2019}
Ewelina Obrzud, Monica Rainer, Avet Harutyunyan, Miles~H. Anderson, Junqiu Liu,
  Michael Geiselmann, Bruno Chazelas, Stefan Kundermann, Steve Lecomte, Massimo
  Cecconi, Adriano Ghedina, Emilio Molinari, Francesco Pepe, François Wildi,
  François Bouchy, Tobias~J. Kippenberg, and Tobias Herr.
\newblock A microphotonic astrocomb.
\newblock {\em Nature Photonics}, 13(1):31--35, January 2019.

\bibitem{parra-rivas_effects_2014}
P.~Parra-Rivas, D.~Gomila, M.~A. Matías, P.~Colet, and L.~Gelens.
\newblock Effects of inhomogeneities and drift on the dynamics of temporal
  solitons in fiber cavities and microresonators.
\newblock {\em Optics Express}, 22(25):30943--30954, December 2014.

\bibitem{jang_all-optical_2016}
Jae~K. Jang, Miro Erkintalo, Jochen Schröder, Benjamin~J. Eggleton, Stuart~G.
  Murdoch, and Stéphane Coen.
\newblock All-optical buffer based on temporal cavity solitons operating at 10
  {Gb/s}.
\newblock {\em Optics Letters}, 41(19):4526--4529, October 2016.

\bibitem{erkintalo_phase_2021}
Miro Erkintalo, Stuart~G. Murdoch, and Stéphane Coen.
\newblock Phase and intensity control of dissipative {Kerr} cavity solitons.
\newblock {\em Journal of the Royal Society of New Zealand}, 0(0):1--19, March
  2021.

\bibitem{jang_temporal_2015}
Jae~K. Jang, Miro Erkintalo, Stéphane Coen, and Stuart~G. Murdoch.
\newblock Temporal tweezing of light through the trapping and manipulation of
  temporal cavity solitons.
\newblock {\em Nature Communications}, 6(1):1--7, June 2015.

\bibitem{jang_writing_2015}
Jae~K. Jang, Miro Erkintalo, Stuart~G. Murdoch, and Stéphane Coen.
\newblock Writing and erasing of temporal cavity solitons by direct phase
  modulation of the cavity driving field.
\newblock {\em Optics Letters}, 40(20):4755--4758, October 2015.

\bibitem{cole_kerr-microresonator_2018}
Daniel~C. Cole, Jordan~R. Stone, Miro Erkintalo, Ki~Youl Yang, Xu~Yi, Kerry~J.
  Vahala, and Scott~B. Papp.
\newblock Kerr-microresonator solitons from a chirped background.
\newblock {\em Optica}, 5(10):1304--1310, October 2018.

\bibitem{weng_microresonator_2022}
Wenle Weng, Jijun He, Aleksandra Kaszubowska-Anandarajah, Prince~M.
  Anandarajah, and Tobias~J. Kippenberg.
\newblock Microresonator {Dissipative} {Kerr} {Solitons} {Synchronized} to an
  {Optoelectronic} {Oscillator}.
\newblock {\em Physical Review Applied}, 17(2):024030, February 2022.

\bibitem{obrzud_temporal_2017}
Ewelina Obrzud, Steve Lecomte, and Tobias Herr.
\newblock Temporal solitons in microresonators driven by optical pulses.
\newblock {\em Nature Photonics}, 11(9):600--607, September 2017.

\bibitem{wang_addressing_2018}
Yadong Wang, Bruno Garbin, François Leo, Stéphane Coen, Miro Erkintalo, and
  Stuart~G. Murdoch.
\newblock Addressing temporal {Kerr} cavity solitons with a single pulse of
  intensity modulation.
\newblock {\em Optics Letters}, 43(13):3192--3195, July 2018.

\bibitem{nielsen_coexistence_2019}
Alexander~U. Nielsen, Bruno Garbin, Stéphane Coen, Stuart~G. Murdoch, and Miro
  Erkintalo.
\newblock Coexistence and {Interactions} between {Nonlinear} {States} with
  {Different} {Polarizations} in a {Monochromatically} {Driven} {Passive}
  {Kerr} {Resonator}.
\newblock {\em Physical Review Letters}, 123(1):013902, July 2019.

\bibitem{nielsen_engineered_2019}
Alexander~U. Nielsen, Yiqing Xu, Michel Ferré, Marcel~G. Clerc, Stéphane
  Coen, Stuart~G. Murdoch, and Miro Erkintalo.
\newblock Engineered discreteness enables observation and control of
  chimera-like states in a system with local coupling.
\newblock {\em arXiv:1910.11329 [nlin, physics:physics]}, October 2019.

\bibitem{strecker_formation_2002}
Kevin~E. Strecker, Guthrie~B. Partridge, Andrew~G. Truscott, and Randall~G.
  Hulet.
\newblock Formation and propagation of matter-wave soliton trains.
\newblock {\em Nature}, 417(6885):150--153, May 2002.

\bibitem{salerno_long-living_2008}
M~Salerno, VV~Konotop, and Yu~V Bludov.
\newblock Long-living {Bloch} oscillations of matter waves in periodic
  potentials.
\newblock {\em Physical review letters}, 101(3):030405, 2008.

\bibitem{michinel_coherent_2012}
Humberto Michinel, \'Angel Paredes, María~M. Valado, and David Feijoo.
\newblock Coherent emission of atomic soliton pairs by {Feshbach}-resonance
  tuning.
\newblock {\em Physical Review A}, 86(1):013620, July 2012.

\bibitem{moura_nonlinear_1994}
M.~A.~de Moura.
\newblock Nonlinear {Schrodinger} solitons in the presence of an external
  potential.
\newblock {\em Journal of Physics A: Mathematical and General}, 27(21):7157,
  November 1994.

\bibitem{serkin_nonautonomous_2007}
V.~N. Serkin, Akira Hasegawa, and T.~L. Belyaeva.
\newblock Nonautonomous {Solitons} in {External} {Potentials}.
\newblock {\em Physical Review Letters}, 98(7):074102, February 2007.

\bibitem{rozenman_observation_2020}
Georgi~Gary Rozenman, Lev Shemer, and Ady Arie.
\newblock Observation of accelerating solitary wavepackets.
\newblock {\em Physical Review E}, 101(5):050201, May 2020.

\bibitem{kutz_modelocked_2006}
J.~Nathan Kutz.
\newblock Mode‐{Locked} {Soliton} {Lasers}.
\newblock {\em SIAM Review}, 48(4):629--678, January 2006.

\bibitem{tusnin_nonlinear_2020}
Aleksandr~K. Tusnin, Alexey~M. Tikan, and Tobias~J. Kippenberg.
\newblock Nonlinear states and dynamics in a synthetic frequency dimension.
\newblock {\em Physical Review A}, 102(2):023518, August 2020.

\bibitem{englebert_bloch_2023}
Nicolas Englebert, Nathan Goldman, Miro Erkintalo, Nader Mostaan, Simon-Pierre
  Gorza, François Leo, and Julien Fatome.
\newblock Bloch oscillations of coherently driven dissipative solitons in a
  synthetic dimension.
\newblock {\em Nature Physics}, 19(7):1014--1021, July 2023.

\bibitem{he_high-speed_2023}
Yang He, Raymond Lopez-Rios, Usman~A. Javid, Jingwei Ling, Mingxiao Li, Shixin
  Xue, Kerry Vahala, and Qiang Lin.
\newblock High-speed tunable microwave-rate soliton microcomb.
\newblock {\em Nature Communications}, 14(1):3467, June 2023.

\bibitem{sun_dissipative_2022}
Yifan Sun, Pedro Parra-Rivas, Mario Ferraro, Fabio Mangini, Mario Zitelli,
  Raphaël Jauberteau, Francesco~Rinaldo Talenti, and Stefan Wabnitz.
\newblock Dissipative {Kerr} solitons, breathers, and chimera states in
  coherently driven passive cavities with parabolic potential.
\newblock {\em Optics Letters}, 47(24):6353--6356, December 2022.

\bibitem{wang_stimulated_2018}
Yadong Wang, Miles Anderson, Stéphane Coen, Stuart~G. Murdoch, and Miro
  Erkintalo.
\newblock Stimulated {Raman} {Scattering} {Imposes} {Fundamental} {Limits} to
  the {Duration} and {Bandwidth} of {Temporal} {Cavity} {Solitons}.
\newblock {\em Physical Review Letters}, 120(5):053902, January 2018.

\bibitem{lugiato_spatial_1987}
L.~A. Lugiato and R.~Lefever.
\newblock Spatial {Dissipative} {Structures} in {Passive} {Optical} {Systems}.
\newblock {\em Physical Review Letters}, 58(21):2209--2211, May 1987.

\bibitem{haelterman_dissipative_1992}
M.~Haelterman, S.~Trillo, and S.~Wabnitz.
\newblock Dissipative modulation instability in a nonlinear dispersive ring
  cavity.
\newblock {\em Optics Communications}, 91(5):401--407, August 1992.

\bibitem{coen_universal_2013}
Stéphane Coen and Miro Erkintalo.
\newblock Universal scaling laws of {Kerr} frequency combs.
\newblock {\em Optics Letters}, 38(11):1790--1792, June 2013.

\bibitem{SuppMat}
See {Supplemental} {Material} for additional information on the normalization
  of the mean-field and the reduced models, the limit to which the {CS} raman
  self-frequency shift can be compensated, {Bloch} oscillations of the {Kelly}
  sidebands, the comparison with external phase modulations, and the
  experimental setup, and additional results on the maximum experimental {CS}
  frequency shift and on the {CS} dynamics over one cavity round-trip.

\bibitem{rogel-salazar_grosspitaevskii_2013}
J.~Rogel-Salazar.
\newblock The {Gross}–{Pitaevskii} equation and {Bose}–{Einstein}
  condensates.
\newblock {\em European Journal of Physics}, 34(2):247, January 2013.

\bibitem{matsko_timing_2013}
Andrey~B. Matsko and Lute Maleki.
\newblock On timing jitter of mode locked {Kerr} frequency combs.
\newblock {\em Optics Express}, 21(23):28862--28876, November 2013.

\bibitem{yi_theory_2016}
Xu~Yi, Qi-Fan Yang, Ki~Youl Yang, and Kerry Vahala.
\newblock Theory and measurement of the soliton self-frequency shift and
  efficiency in optical microcavities.
\newblock {\em Optics Letters}, 41(15):3419--3422, August 2016.

\bibitem{veltz_bifurcationkitjl_2020}
Romain Veltz.
\newblock {BifurcationKit}.jl, July 2020.

\bibitem{longhi_variational_1997}
S.~Longhi, G.~Steinmeyer, and W.~S. Wong.
\newblock Variational approach to pulse propagation in parametrically amplified
  optical systems.
\newblock {\em JOSA B}, 14(8):2167--2173, August 1997.

\bibitem{englebert_temporal_2021}
Nicolas Englebert, Carlos Mas~Arabí, Pedro Parra-Rivas, Simon-Pierre Gorza,
  and François Leo.
\newblock Temporal solitons in a coherently driven active resonator.
\newblock {\em Nature Photonics}, 15(7):536--541, 2021.

\bibitem{englebert_high_2023}
Nicolas Englebert, Carlos~Mas Arabí, Simon-Pierre Gorza, and François Leo.
\newblock High peak-to-background-ratio solitons in a coherently driven active
  fiber cavity.
\newblock {\em APL Photonics}, 8(12):120802, December 2023.

\bibitem{karpov_raman_2016}
Maxim Karpov, Hairun Guo, Arne Kordts, Victor Brasch, Martin~H.P. Pfeiffer,
  Michail Zervas, Michael Geiselmann, and Tobias~J. Kippenberg.
\newblock Raman {Self}-{Frequency} {Shift} of {Dissipative} {Kerr} {Solitons}
  in an {Optical} {Microresonator}.
\newblock {\em Physical Review Letters}, 116(10):103902, March 2016.

\bibitem{lei_optical_2022}
Fuchuan Lei, Zhichao Ye, Óskar~B. Helgason, Attila Fülöp, Marcello Girardi,
  and Victor Torres-Company.
\newblock Optical linewidth of soliton microcombs.
\newblock {\em Nature Communications}, 13(1):3161, June 2022.

\bibitem{atieh_measuring_1999}
A.K. Atieh, P.~Myslinski, J.~Chrostowski, and P.~Galko.
\newblock Measuring the {Raman} time constant ({T}/sub {R}/) for soliton pulses
  in standard single-mode fiber.
\newblock {\em Journal of Lightwave Technology}, 17(2):216--221, February 1999.

\bibitem{luo_resonant_2015}
Kathy Luo, Yiqing Xu, Miro Erkintalo, and Stuart~G. Murdoch.
\newblock Resonant radiation in synchronously pumped passive {Kerr} cavities.
\newblock {\em Optics Letters}, 40(3):427--430, February 2015.

\bibitem{chen_real-time_2021}
Hao Chen, NingNing Yang, Chengzhi Qin, Wenwan Li, Bing Wang, Tianwen Han, Chi
  Zhang, Weiwei Liu, Kai Wang, Hua Long, Xinliang Zhang, and Peixiang Lu.
\newblock Real-time observation of frequency {Bloch} oscillations with fibre
  loop modulation.
\newblock {\em Light: Science \& Applications}, 10(1):48, March 2021.

\bibitem{pedaci_positioning_2006}
F.~Pedaci, P.~Genevet, S.~Barland, M.~Giudici, and J.~R. Tredicce.
\newblock Positioning cavity solitons with a phase mask.
\newblock {\em Applied Physics Letters}, 89(22):221111, November 2006.

\bibitem{yao_dual_1997}
X.~Steve Yao and Lute Maleki.
\newblock Dual microwave and optical oscillator.
\newblock {\em Optics Letters}, 22(24):1867--1869, December 1997.

\bibitem{yu_ultralow-noise_2005}
Nan Yu, Ertan Salik, and Lute Maleki.
\newblock Ultralow-noise mode-locked laser with coupled optoelectronic
  oscillator configuration.
\newblock {\em Optics Letters}, 30(10):1231--1233, May 2005.

\bibitem{englebert_parametrically_2021}
Nicolas Englebert, Francesco De~Lucia, Pedro Parra-Rivas, Carlos~Mas Arabí,
  Pier-John Sazio, Simon-Pierre Gorza, and François Leo.
\newblock Parametrically driven {Kerr} cavity solitons.
\newblock {\em Nature Photonics}, 15(11):857--861, November 2021.

\bibitem{moille_parametrically_2024}
Grégory Moille, Miriam Leonhardt, David Paligora, Nicolas Englebert, François
  Leo, Julien Fatome, Kartik Srinivasan, and Miro Erkintalo.
\newblock Parametrically driven pure-{Kerr} temporal solitons in a
  chip-integrated microcavity.
\newblock {\em Nature Photonics}, 18(6):617--624, June 2024.

\bibitem{rozenman_amplitude_2019}
Georgi~Gary Rozenman, Matthias Zimmermann, Maxim~A. Efremov, Wolfgang~P.
  Schleich, Lev Shemer, and Ady Arie.
\newblock Amplitude and {Phase} of {Wave} {Packets} in a {Linear} {Potential}.
\newblock {\em Physical Review Letters}, 122(12):124302, March 2019.

\end{thebibliography}
\bibliographystyle{unsrt}

\foreach \x in {1,...,8}
{%
\clearpage
\includepdf[pages={\x}]{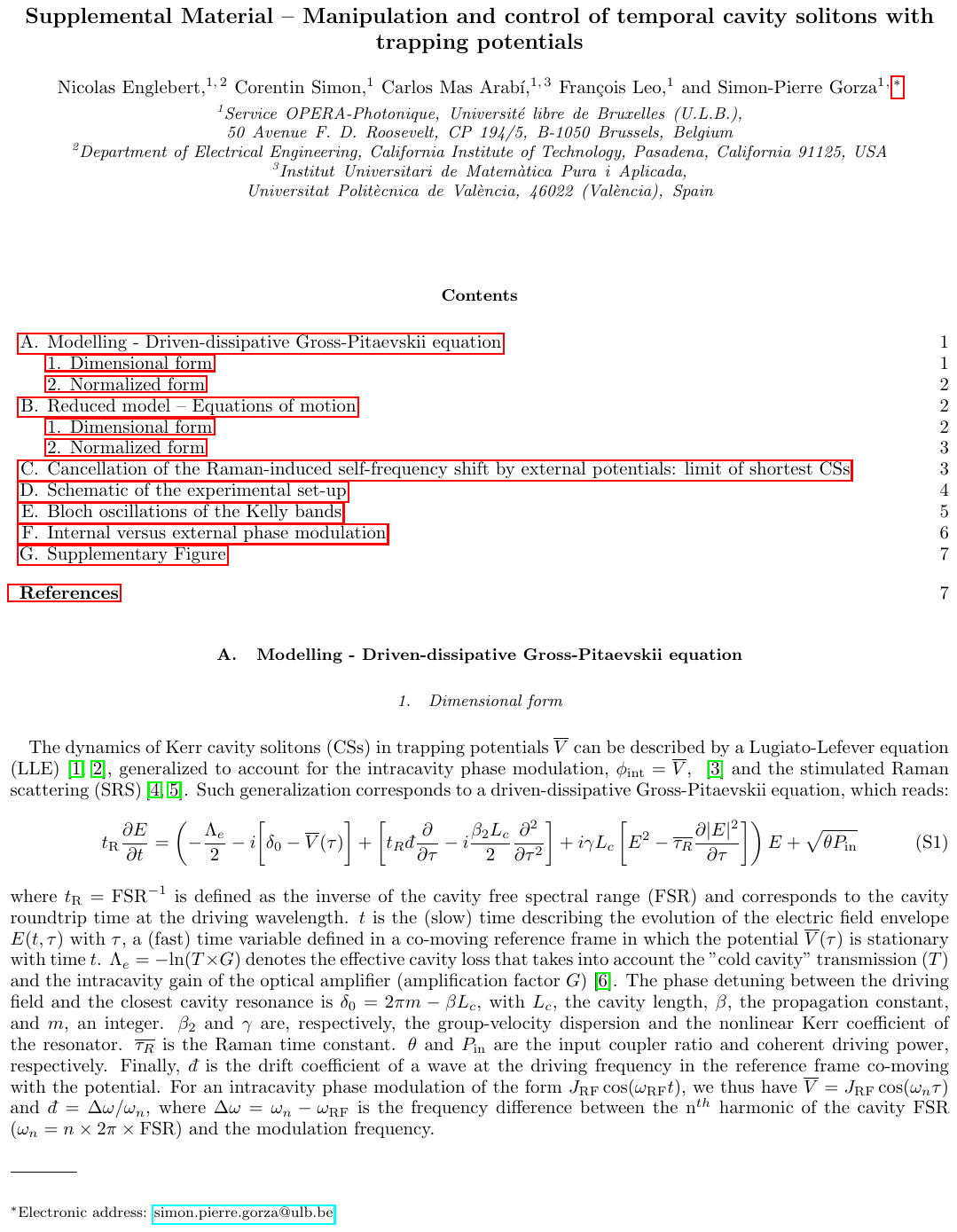} 
}

\end{document}